\documentclass[pre,preprint,showpacs]{revtex4}
\usepackage{epsfig}

\begin{document}

\title{Canonical variables for steep planar water waves over nonuniform bed}
\author{V. P. Ruban}
\email{ruban@itp.ac.ru}
\affiliation{Landau Institute for Theoretical Physics,
2 Kosygin Street, 119334 Moscow, Russia} 

\date{\today}

\begin{abstract}
An explicit expression in terms of canonical variables is obtained 
for the Hamiltonian functional determining the fully nonlinear dynamics of
two-dimensional potential flows of an ideal fluid with a free surface
over an arbitrary nonuniform depth. The canonically conjugate variables
are derived from the previously developed non-canonical conformal
description of water waves over a strongly undulating bottom
[V. P. Ruban, Phys. Rev. E {\bf 70}, 066302 (2004)]. Also an alternative 
approach to the problem is discussed, which gives weakly nonlinear 
Hamiltonian models of different orders.
\end{abstract}




\maketitle


\section{Introduction}

For many phenomena related to the science of ocean waves, 
it is necessary to study fully nonlinear regimes in the dynamics 
of free water surface. For instance, rogue waves are highly nonlinear 
structures attracting much interest over past years \cite{Kharif-Pelinovsky}. 
Breaking waves on a beach give another example of nonlinear 
water wave dynamics. 
For some problems the simplified mathematical formulation can be used, 
when the fluid is treated as inviscid and incompressible, 
and only potential flows with a constant pressure above the free surface 
are considered. Such class of ideal flows is known to possess the Hamiltonian 
structure \cite{Z68}, where the pair of canonically conjugate 
functions is formed, for example,  by the vertical displacement 
$\eta({\bf r},t)$ of the free surface from the mean horizontal level 
and by the boundary distribution $\psi({\bf r},t)$ of the velocity potential, 
with ${\bf r}$ being the position in the horizontal plane and $t$ 
being the time. Unfortunately,
the variables $\eta$ and $\psi$ are not suitable for highly nonlinear 
waves, since there is no explicit expression for the Hamiltonian functional 
in terms of $\eta$ and $\psi$, valid for arbitrary wave steepness. 
For two-dimensional (2D) flows (in $xy$-plane) this difficulty can be avoided, 
and exact and compact equations of motion are possible 
with the help of so called conformal variables originally introduced
by L.V. Ovsyannikov in early 1970's \cite{Ovs73,Ovs74}. 
The conformal description was combined with a variational 
formalism in 1990's by V. E. Zakharov and 
co-workers, for the infinite depth \cite{DKSZ96} 
and for the case of flat horizontal bed \cite{DZK96}. 
Subsequent development of this method was done in 
Refs.\cite{D2001,DLZ95,L97,CC1999}.
The general idea of the approach is to use a 
time-dependent conformal mapping $x+iy=z(w,t)$ which ``straightens'' 
the flow domain, with an analytic function $z(w,t)$ 
depending on a complex variable $w=u+iv$.
The free surface and the bottom correspond to some fixed values $v_s$ and $v_b$ 
of the curvilinear coordinate $v$ (in the case of infinite depth $v_s=0$ and
$v_b=-\infty$, while in the case of finite depth without loss of generality
one can take $v_s=1$ and $v_b=0$), 
and  thus the conformally invariant 2D Laplace 
equation for the velocity potential is easily solved. Basic unknown dynamical 
quantities, such as the boundary value  $\psi$ of the velocity potential
and the vertical coordinate $Y$ along the free surface, 
in this description are functions of $u$ and $t$. Explicit expressions 
for time derivatives $\psi_t(u,t)$ and $Y_t(u,t)$ can be derived 
from variational Euler-Lagrange equations with the help of
a linear integral relation between $Y(u,t)$ and $X(u,t)$,
taking place due to the analyticity \cite{DKSZ96,DZK96}.

Non-trivial bottom topography is known to affect wave motion.
In order to include into the fully nonlinear model that influence, 
the conformal description was generalized by the present author to the 
case of static nonuniform bed \cite{R2004PRE}, for time-dependent bottom 
boundary \cite{R2005PLA}, and for planar flows with constant vorticity 
over nonuniform bed \cite{R2008PRE}. 
Even for weakly nonlinear regimes, conformal coordinates in the problems 
with bottom topography are more preferable than the Cartesian coordinates,
though in that case a somewhat different approach is employed
(see, for example, Refs.\cite{N2003,AN2004,R2004PRE}, and Sec.III below).
A step towards using conformal variables for three-dimensional (3D) highly 
nonlinear potential flows with a free surface was done  
in the paper \cite {R2005PRE}, where for long-crested waves
an asymptotic expansion of the Hamiltonian functional was suggested 
in the powers of a small parameter corresponding to the 
squared ratio of a typical wave length $\lambda_0$ to a long typical distance 
$l_q$ along wave crests stretching in the transversal horizontal direction $q$.

Equations of motion in the conformal description contain integral operators 
which are diagonal in the Fourier representation. 
Therefore in the numerical simulations a Fast Fourier Transform (FFT) 
can be used (see, for example, the web site \cite{fftw3} about a modern 
and efficient FFT library). Important numerical results concerning  planar 
rogue waves on infinitely deep water have been obtained 
by this method in Refs.\cite{ZDV2002,DZ2005,ZDP2006,DZ2008}. 
Nonlinear waves over inhomogeneous static and time-dependent beds were 
simulated in the conformal variables in Refs.\cite{R2004PRE,R2005PLA}. 
The most recent application of this method can be found in 
Refs.\cite{R2008PRE-2,R2008PRE-3} where the phenomenon of so called 
water-wave gap solitons over spatially periodic beds was studied.
The weakly 3D conformal theory \cite {R2005PRE} has been recently successfully 
applied to simulate long-crested freak waves at infinitely deep water
\cite{RD2005PRE,R2006PRE,R2007PRL,R2009PRE}.

It should be noted that in most cited papers non-canonical conformal variables 
were used, as they are quite sufficient for numerical simulations. 
The exceptions are the analytical works \cite{DLZ95,L97}, where a pair of 
canonical conformal variables has been employed, however in the case of 
infinite depth only. For uneven beds, the canonical pair was 
not derived until now. In the present work, this gap in the theory will 
be filled and canonical conformal variables together with an explicit 
Hamiltonian will be suggested for 2D ideal potential flows over 
an arbitrary nonuniform bottom profile (in Sec.II). 
But we will see the obtained exact Hamiltonian contains 
a strongly non-local linear operator which at long scales is similar 
to the integrating operator $\partial_u^{-1}$. The presence of such operator 
makes the exact canonical model somewhat difficult for analysis.
Therefore an alternative approach how to introduce 
canonical variables for waves over strongly undulating bottom will be also
considered (in Sec.III), which is free of the above mentioned technical 
difficulty. Unfortunately, the alternative approach does not allow us to obtain 
an exact Hamiltonian, but approximations of different orders 
for weakly nonlinear waves are possible. In this paper, the
calculations will be made up to the fifth order.

\section{Exact Hamiltonian for waves over nonuniform depth}

It is appropriate to mention here that if the given bottom boundary 
is non-flat, then the conformal mapping $z(w,t)$ 
can be represented as the composition $z(w,t)=Z(\zeta(w,t),t)$, 
where $Z(\zeta,t)$ is a known analytic
function [$Z(\zeta)$ does not depend on time if the bed is static,
which case is assumed in this work]. 
The intermediate unknown analytic function $\zeta(w,t)$ takes purely real 
values $\gamma(u,t)$ at the real axis: $\zeta(u+0i,t)=\gamma(u,t)$, 
and $\gamma_u(u,t) > 0$. 
As a result, the bed profile is given in the parametric form 
$X^{[b]}+iY^{[b]}=Z(\gamma)$, where $-\infty<\gamma<+\infty$,
while the shape of free surface is determined parametrically by the formula
\begin{equation}
X^{[s]}+iY^{[s]}=Z(\xi(u,t)),\qquad \xi(u,t)=\zeta(u+i,t).
\end{equation}
Since $\zeta(u+i,t)$ is the analytic continuation of a real function 
$\gamma(u,t)$
from the real axis, a linear integral relation takes place between 
the real and the imaginary parts of function $\xi(u,t)$:
\begin{equation}
\xi(u,t)=e^{-\hat k}\gamma(u,t)=(1+i\hat R)\rho(u,t), 
\end{equation}
where $\hat k=-i\hat \partial_u$ is the differential operator, 
$e^{-\hat k}$ is the operator making the analytic continuation, 
$\rho(u,t)=[\cosh \hat k]\gamma(u,t)$ is another unknown real function, 
and $\hat R=i[\tanh \hat k]$ 
is a linear antisymmetric operator which is diagonal in the 
Fourier representation:
\begin{equation}
\hat R\rho(u,t)=\int i[\tanh k] \,\rho_k(t)e^{iku}\frac{dk}{2\pi},
\end{equation}
with $\rho_k(t)=\int \rho(u,t)e^{-iku}du$ being the Fourier transform of the 
function $\rho(u,t)$. The inverse operator is 
$\hat T=\hat R^{-1}=-i[\coth \hat k]$. For long waves, when $|k|\ll 1$, 
operator $\hat R$ is similar to $\hat \partial_u$, while
$\hat T$ is similar to integrating operator $\hat \partial_u^{-1}$.
Explicit formulas for the kernels of operators $\hat R$ and $\hat T$ 
in $u$-representation are given, for instance, in Ref.\cite{R2004PRE}.

A simple way to obtain the required canonical pair is the following. 
We first recall the expression for the Lagrangian functional 
describing water waves over a static nonuniform bed in terms of 
the conformal variables (see Ref.\cite{R2004PRE}),
\begin{equation}\label{Lagrangian_rho_psi}
{\cal L}=\int \psi|Z'(\xi)|^2(
\rho_u \hat R\rho_t-\rho_t\hat R\rho_u) \,du -{\cal H}.
\end{equation}
Here $Z'(\xi)\equiv dZ/d\xi$,
and the Hamiltonian ${\cal H}$ is the total energy of the system ---
the kinetic energy plus the potential energy in the vertical 
gravitational field $g$ (surface-tension effects can be also included, 
but we do not consider them in this paper),
\begin{equation}\label{Hamiltonian_rho_psi}
{\cal H}=\frac{1}{2}\int\psi\hat K \psi \,du 
+\frac{g}{2}\int\left[\mbox{Im }Z(\xi)\right]^2
\mbox{Re}\left[Z'(\xi)\xi_u\right]du,
\end{equation}
where $\hat K\equiv \hat k \tanh \hat k$ is a Hermitian operator. 

Let us now choose function $\rho(u,t)$ as the generalized canonical coordinate. 
Using the antisymmetric property of operator $\hat R$,
we rewrite Eq.(\ref{Lagrangian_rho_psi}) as follows,
\begin{equation}
{\cal L}=\int \left\{-\hat R\left[\psi|Z'(\xi)|^2 \rho_u\right]
-\psi|Z'(\xi)|^2 \hat R \rho_u \right\}\rho_t \,du -{\cal H}.
\end{equation}
It is clear that the Lagrangian will take the  canonical form
${\cal L}=\int \mu\rho_t \,du -{\cal H}$
if we define  the canonical momentum $\mu$ by the relation written below
\begin{equation}\label{mu_definition}
\mu=-\hat R\left[\psi|Z'(\xi)|^2 \rho_u\right]
-\psi|Z'(\xi)|^2 \hat R \rho_u.
\end{equation}
Using this equation, it is now necessary to express $\psi$ through 
the canonical variables $\mu$ and $\rho$ and then substitute 
the result into Eq.(\ref{Hamiltonian_rho_psi}). Generally speaking, 
this is not a trivial task in view of the presence of integral operator 
$\hat R$. Fortunately, the coefficients of the equation are very special, 
and therefore the solution can be found in exact form.
To solve the integral equation (\ref{mu_definition}) with respect to $\psi$,
we rewrite it in the form
\begin{equation}
\mbox{Re}\left\{[1+i\hat R]\big(-i\psi|Z'(\xi)|^2\xi_u +\mu\big)\right\}=0.
\end{equation}
The above equation allows us to conclude that
\begin{equation}\label{f}
-i\psi|Z'(\xi)|^2\xi_u +\mu=-i(1-i\hat R)f,
\end{equation} 
where $f$ is some real function. Since $\bar\xi_u=(1-i\hat R)\rho_u$ 
(the overline means the complex conjugate quantity), 
we can multiply Eq.(\ref{f}) by $i\bar\xi_u$ and use a general formula
\begin{equation}
[(1-i\hat R)f_1][(1-i\hat R)f_2]=(1-i\hat R)f_3, 
\end{equation}
where $f_1$, $f_2$, and $f_3$  are real functions, and 
$f_3=f_1 f_2-(\hat R f_1) (\hat R f_2)$.
As the result, we have 
\begin{equation}\label{tilde_f}
\psi|Z'(\xi)\xi_u|^2 +i\mu\bar\xi_u=(1-i\hat R)\tilde f,
\end{equation} 
where $\tilde f$ is another real function. Taking the imaginary part
of Eq.(\ref{tilde_f}),  we obtain $\mu \rho_u=-\hat R\tilde f$, and thus
$\tilde f=-\hat T(\mu\rho_u)$. 
Then the real part gives us the required formula
\begin{equation}\label{psi_PQ}
\psi=\frac{-\hat T(\mu\,\rho_u)-\mu\,\hat R\rho_u}
{|\hat\partial_u Z([1+i\hat R]\rho)|^2}.
\end{equation} 
Now we are able to write explicit and exact expression for the Hamiltonian 
functional in terms of the canonical variables $\rho$ and  $\mu$,
\begin{eqnarray}
{\cal H}&\!=\!&\frac{1}{2}\!\int\! 
\Bigg[\frac{\hat T(\mu\,\rho_u)+\mu\,\hat R\rho_u}
{|\hat\partial_u Z([1+i\hat R]\rho)|^2}\Bigg]
\hat K \Bigg[\frac{\hat T(\mu\,\rho_u)+\mu\,\hat R\rho_u}
{|\hat\partial_u Z([1+i\hat R]\rho)|^2}\Bigg]du \nonumber\\
&\!+\!&\frac{g}{2}\!\int \!\big[\mbox{Im }Z([1+i\hat R]\rho)\big]^2
\mbox{Re}\big[\hat\partial_u Z([1+i\hat R]\rho)\big]du,
\label{H_rho_mu}
\end{eqnarray}
which is the central result of this paper. The obtained formula is rather
cumbersome, and it is unclear yet if some canonical transformation will 
be able to reduce this Hamiltonian to a simpler form.
Further serious work on the problem how to simplify the Hamiltonian 
(\ref{H_rho_mu}) is needed.

The corresponding canonical equations of motion are 
$\rho_t=\delta{\cal H}/\delta\mu$ and $-\mu_t=\delta{\cal H}/\delta\rho$,
where the variational derivatives should be calculated from 
Eq.(\ref{H_rho_mu}) according to the well-established general rules. 
Let us introduce a short-hand notation
\begin{equation}\label{N}
N\equiv|\hat\partial_u Z(\xi)|^{-2}
\hat K \Bigg[\frac{\hat T(\mu\,\rho_u)+\mu\,\hat R\rho_u}
{|\hat\partial_u Z(\xi)|^2}\Bigg].
\end{equation}
Then the variational derivatives are
\begin{equation}
\frac{\delta {\cal H}}{\delta\mu}=(\hat R\rho_u) N-\rho_u\hat T N,
\label{dH_dmu}
\end{equation}
\begin{eqnarray}
\frac{\delta {\cal H}}{\delta\rho}&=&
\hat\partial_u[\hat R( \mu N) +\mu\hat T N]\nonumber\\
&-&2\mbox{Re}\left[(1-i\hat R)
\{[\hat T(\mu\,\rho_u)+\mu\,\hat R\rho_u]NZ''(\xi)/Z'(\xi)\}\right]\nonumber\\
&+&2\mbox{Re}\left[(1-i\hat R)
\hat\partial_u \{[\hat T(\mu\,\rho_u)+\mu\,\hat R\rho_u]N/\xi_u\}\right]
\nonumber\\
&+&g\,\mbox{Im}\left[(1-i\hat R)
\{\mbox{Im}(Z(\xi)) |Z'(\xi)|^2\bar\xi_u\}\right].
\label{dH_drho}
\end{eqnarray}

\section{Alternative approach}

An essential difficulty of the derived fully nonlinear canonical model
is the presence of strongly non-local operator $\hat T$. Therefore
it makes sense to consider alternative approaches to the problem of 
canonical variables for waves over uneven bed. Below we shall focus
on the approximate method suggested in Ref.\cite{R2004PRE},
where curvilinear conformal coordinates $u$ and $v$ are static. 
In this approach, the conformal mapping ``straightens'' the bottom boundary 
but not the free surface, in contrast to the exact description used 
in the previous section. In Ref.\cite{R2004PRE},
the corresponding Hamiltonian functional was derived up to the third order
in canonical variables. Here we develop this method more systematically 
and calculate the fourth- and fifth-order approximations. 

We now introduce (static) curvilinear coordinates $u(x,y)$ and $v(x,y)$, where
real function $v(x,y)$ obeys the Laplace equation $v_{xx}+v_{yy}=0$ with the 
boundary conditions $v=-1$ at the given arbitrary nonuniform bottom and $v=0$ 
at $y=0$. Function $u(x,y)$ is taken harmonically conjugate for $v(x,y)$.
The inverse conformal mapping $x+iy=z(u+iv)$ satisfies the condition
$\mbox{Im }z(u+0i)=0$, so a real function $x(u)=z(u+0i)$
parametrizes the unperturbed free surface $y=0$. The bed profile 
is given in the parametric form $X^{[b]}+iY^{[b]}=z(u-i)$, while the shape 
of the free boundary is determined through an unknown real function $V(u,t)$ 
by the complex equality
\begin{equation}\label{Z_V}
X^{[s]}+iY^{[s]}=z\big(u+iV(u,t)\big).
\end{equation}
The free-boundary value of the velocity potential is 
$\psi(u,t)=\varphi(u,V(u,t),t)$,
where the potential $\varphi(u,v,t)$ obeys 2D Laplace equation 
$\varphi_{uu}+\varphi_{vv}=0$ in the flow domain $-1\le v\le V(u,t)$, 
with the bottom boundary condition $\varphi_v(u,-1,t)=0$. Consequently, 
a general form of the Lagrangian functional in this description 
is the following,
\begin{equation}\label{L_V_psi}
{\cal L}=\int \psi|z'(u+iV)|^2V_t du -{\cal H}\{V,\psi\}.
\end{equation}
The Hamiltonian functional 
${\cal H}\{V,\psi\}={\cal K}\{V,\psi\}+{\cal P}\{V\}$, where 
${\cal K}\{V,\psi\}$ is the kinetic energy and ${\cal P}\{V\}$ 
is the potential energy. Functional ${\cal P}\{V\}$ is relatively simple,
\begin{eqnarray}
{\cal P}&=&\frac{g}{2}\int[\mbox{Im }z(u+iV)]^2
\mbox{Re}[\hat\partial_uz(u+iV)]du\nonumber\\
&=&\frac{g}{2}\int 
\left\{V^2x'^3+V^4\left[\frac{(x'^2x'')'}{4}-\frac{5}{6}x'^2x'''\right]\right\}
du\nonumber\\
&&\quad+\,{\cal O}(V^6),
\label{P_V_psi}
\end{eqnarray}
with $x'=dx(u)/du$, $x''=d^2x(u)/du^2$ and so on. As to
the kinetic energy ${\cal K}\{V,\psi\}$, it cannot be represented 
in an exact form, but its expansion 
${\cal K}={\cal K}^{[2]}+{\cal K}^{[3]}+{\cal K}^{[4]}+\cdots$
in powers of (supposedly small) functions $V$ and $\psi$ is possible.
We note that the equations determining the 2D velocity potential 
$\varphi(u,v,t)$ 
are formally identical to the equations for the velocity potential 
$\varphi(x,y,t)$ in Cartesian coordinates over a straight horizontal bottom
at $y=-1$.
The kinetic energy is determined by the same integral in both cases,
\begin{equation}\label{K_def}
{\cal K}\{V,\psi\}=\frac{1}{2}\int du\int_{-1}^{V(u)}(\varphi_u^2+\varphi_v^2)dv.
\end{equation}
In view of this equivalence, the expansion of ${\cal K}\{V,\psi\}$
is calculated in the same standard manner as 
for weakly nonlinear waves over flat bottom in Cartesian coordinates 
(compare to expansion of ${\cal K}\{\eta,\psi\}$ in Appendix B 
of Ref.\cite{R2008PRE-3}):
\begin{eqnarray}
{\cal K}^{[2]}&=&\frac{1}{2}\int\psi\hat K\psi du, 
\label{K2_V_psi}\\
{\cal K}^{[3]}&=&
\frac{1}{2}\int V[(\psi_u)^2-(\hat K\psi)^2]du,
\label{K3_V_psi}\\
{\cal K}^{[4]}&=&\frac{1}{2}\int\left[
\psi\hat K V \hat K V\hat K\psi 
+V^2(\hat K\psi)\psi_{uu} \right] du.
\label{K4_V_psi}
\end{eqnarray}
\begin{eqnarray}\label{K5_V_psi}
{\cal K}^{[5]}&=&\frac{1}{2}\int\Big[ 
\frac{V^3}{6}(\hat K\psi)\hat K\psi_{uu}
- \psi \hat K V \hat K V \hat K V \hat K\psi\nonumber\\
&&\qquad-\frac{V^3}{3}(\psi_{uu})^2
-V^2(\hat K V \hat K\psi)\psi_{uu}\nonumber\\
&&\qquad-\frac{V^2}{2}(\hat K\psi)\partial_u^2(V \hat K\psi)
\Big] du.
\end{eqnarray}
As a consequence,
the bed nonuniformity does not affect ${\cal K}\{V,\psi\}$ in any order. 
This is an essential technical advantage of the conformal coordinates 
in this problem.

The form of Lagrangian (\ref{L_V_psi}) allows us to treat $\psi(u)$ 
as the canonical momentum if the corresponding canonical coordinate 
$\chi(u)$ is defined in the following way,
\begin{equation}\label{chi_definition} 
\chi(u)=\int_0^V |z'(u+iv)|^2dv.
\end{equation}
Since we consider small values $V$, the above equation gives us an expansion
of $\chi$ in odd powers of $V$,
\begin{equation}\label{chi_approx} 
\chi= x'^2 V+\big[(x'')^2-x'x'''\big]V^3/3 +{\cal O}(V^5).
\end{equation}
From here we can express the non-canonical variable $V$ through the
canonical variable $\chi$:
\begin{equation}\label{V_approx} 
V=\frac{\chi}{x'^2}\left[1+\left(\frac{x'x'''-(x'')^2}{x'^6}\right)
\frac{\chi^2}{3}+{\cal O}(\chi^4)\right].
\end{equation}
Now we substitute the above expression into ${\cal H}\{V,\psi\}$ 
and collect there terms of the same order. 
After simplification, we obtain the Hamiltonian functional 
${\cal H}\{\chi,\psi\}={\cal H}^{[2]}+{\cal H}^{[3]}+\cdots$  
up to the fifth order:
\begin{eqnarray}
{\cal H}^{[2]}&=&\frac{1}{2}\int\left[\psi\hat K\psi +\frac{g}{x'}\chi^2\right]
du, \label{H2_chi_psi}\\
{\cal H}^{[3]}&=&
\frac{1}{2}\int[(\psi_u)^2-(\hat K\psi)^2]\frac{\chi}{x'^2}du,
\label{H3_chi_psi}\\
{\cal H}^{[4]}&=&\frac{1}{2}\int\left[
\psi\hat K \frac{\chi}{x'^2} \hat K \frac{\chi}{x'^2}\hat K\psi 
+\frac{\chi^2}{x'^4}(\hat K\psi)\psi_{uu} \right] du \nonumber\\
&+&\frac{g}{12}\int\left[\frac{1}{2}\frac{x'''}{x'^6}
-\frac{(x'')^2}{x'^7} \right]\chi^4du.
\label{H4_chi_psi}
\end{eqnarray}
\begin{eqnarray}
&&{\cal H}^{[5]}=\frac{1}{6}\int[(\psi_u)^2-(\hat K\psi)^2]
\left[\frac{x'x'''-(x'')^2}{x'^8}\right]\chi^3 du\nonumber\\
&&+\frac{1}{12}\int\Big[ 
\frac{\chi^3}{x'^6}(\hat K\psi)\hat K\psi_{uu}
-6 \psi \hat K \frac{\chi}{x'^2} \hat K  \frac{\chi}{x'^2} \hat K 
 \frac{\chi}{x'^2} \hat K\psi\nonumber\\
&&\qquad-2\frac{\chi^3}{x'^6}(\psi_{uu})^2
-6\frac{\chi^2}{x'^4}(\hat K \frac{\chi}{x'^2} \hat K\psi)\psi_{uu}\nonumber\\
&&\qquad-3\frac{\chi^2}{x'^4}(\hat K\psi)\partial_u^2
(\frac{\chi}{x'^2} \hat K\psi)
\Big] du.
\label{H5_chi_psi}
\end{eqnarray}
We do not write here the corresponding canonical equations of motion ---
though it is a simple exercise to calculate the variational derivatives
$\delta{\cal H}/\delta\chi$ and $\delta{\cal H}/\delta\psi$, but the formulas
are quite long.
This weakly nonlinear model can be applied to a wide variety of problems, 
for instance, to study the dynamics of solitons propagating along a channel 
of variable depth, or the scattering of a wave train by a strong bed 
nonuniformity. The obvious advantages of the model are that it is fully 
dispersive and  there is no principal limitations on the bottom profiles.
However, it is not clear at the moment what is the largest value of wave 
amplitude which is still well described by this approximate Hamiltonian. 
The answer perhaps depends on many details. This question requires further 
investigations.

Weakly dispersive (shallow-water) regime can take place if the depth is 
slowly varying, $|x''|/x'\ll 1$. In that regime
one can reduce the Hamiltonian to a purely local form by
using the long-wave expansion for the integral operator 
$\hat K=-\hat\partial_u^2-\hat\partial_u^4/3+\cdots$ and then apply
some analytical methods.
Otherwise, the  fully dispersive equations of motion can be easily simulated 
on computer with the help of FFT routines, since all linear operators 
are diagonal in Fourier representation.

To conclude, in this paper two different methods have been presented 
how to solve the theoretical problem of canonical Hamiltonian
description of nonlinear planar water waves over nonuniform depth. 
The first method is fully nonlinear, but the corresponding Hamiltonian 
functional contains a strongly nonlocal operator. 
The second way is the further development of the weakly nonlinear 
approach suggested in Ref.\cite{R2004PRE}.

These investigations were supported by RFBR 
(grants 09-01-00631 and 07-01-92165),
by the ``Leading Scientific Schools of Russia'' grant 6885.2010.2,
and by the Program ``Fundamental Problems of Nonlinear Dynamics'' 
from the RAS Presidium.

\end{document}